\documentclass[preprint,prd,showpacs]{revtex4} 
 
\begin{document} 

\title{Effects of Quantum Stress Tensor Fluctuations with
Compact Extra Dimensions}

\author{J. Borgman}
\author{ L.H. Ford} 
 \email[Email: ]{ford@cosmos.phy.tufts.edu} 
 \affiliation{Institute of Cosmology  \\
Department of Physics and Astronomy\\ 
         Tufts University, Medford, MA 02155}

\begin{abstract} 
The effects of compact extra dimensions upon quantum stress tensor
fluctuations are discussed. It is argued that as the compactification 
volume decrease, these fluctuations increase in magnitude. In principle,
this would have the potential to create observable effects, such as 
luminosity fluctuations or angular blurring of distant sources, and
lead to constraints upon Kaluza-Klein theories.  However,
the dependence of the four-dimensional Newton's constant upon the
compactification volume causes the gravitational effects of the stress tensor
fluctuations to be finite in the limit of small volume. Consequently,
no observational constraints upon Kaluza-Klein theories are obtained. 
\end{abstract}

\maketitle

\baselineskip=13pt 

\section{Introduction}

Theories with extra spacetime dimensions were introduced into physics
long ago by Kaluza~\cite{Kaluza} and by Klein~\cite{Klein}. The original
Kaluza-Kaluza theory postulated a fifth spacetime dimension as part
of an attempt to unify gravity and electromagnetism. In order that we
not directly observe the fifth dimension, it is assumed to be compact
with a very small compactification length. Modern versions of Kaluza-Kaluza 
theories come in a variety of forms~\cite{ACF}. Among the higher
dimensional models to attract considerable attention in recent years
are the ``braneworlds'' models~\cite{ADD} and Randall-Sundrum type 
models~\cite{RS99}. The former postulate that all fields except gravity 
are confined to a four-dimensional brane, whereas gravity is free to
propagate in two or more extra dimensions. All of these
higher dimensional models require an explanation as to why the world
we observe appears to have only four spacetime dimensions.

In  Kaluza-Kaluza theories, the extra dimensions are compact, although
they may be either flat or curved. Thus if the size of the extra dimensions
is less than the smallest scale at which we can probe experimentally,
then the extra dimensions should not be seen. If the largest compact
dimension has a size $L$, then an energy of the order of $1/L$ is required
to excite momenta in this dimension. The conventional view is that if
we have performed scattering experiments at an energy scale of $E$
and not seen any effects of extra dimensions, then their sizes are
constrained to be less than about $1/E$. 

However, there is a possible loophole in this reasoning. If there are
quantum fields propagating in all spacetime dimensions, then the
fluctuation effects of these fields will {\it grow}, not diminish,
as the size of the extra dimensions decreases. This can be seen as
a consequence of the uncertainty principle: a quantum system confined
to a smaller volume of configuration space must exhibit increased
fluctuations in the conjugate variables. A simple illustration of this 
comes from the way Casimir energy densities scale; in $d$ spacetime dimensions,
a quantum field confined by boundaries or periodicity on a scale $L$
will have a Casimir energy density of the order of $L^{-d}$. That
Casimir energy creates a potential problem for Kaluza-Kaluza theories has long
been recognized. There are two approaches which have been taken to dealing
with this large energy density. One is to worry only about the projection
of the stress tensor into the four dimensional uncompactified spacetime.
Here Lorentz symmetry requires this part of the stress tensor to be
proportional to the four-dimensional metric tensor, that is, to be of
the form of the cosmological constant. In this case, it can be absorbed
by a cosmological constant renormalization. This is not a completely
satisfactory solution, however, as it does not address the effects of the
large stress components in the compact dimensions. If one wants the
Einstein equations in $d$  dimensions to be satisfied, there is a nontrivial
stabilization problem. Another approach which has been proposed involves
cancelling the Casimir energies of various fields, such as bosons and
fermions~\cite{CW84}.

Even if renormalization or cancellation between different quantum fields
succeed in making the net expectation value of the stress tensor,
$\langle T^{\mu\nu} \rangle$, acceptably small, there is no guarantee
that the fluctuations around the mean value will be small. A cosmological
constant counter term can only shift the mean value, 
$\langle T^{\mu\nu} \rangle$, but has no effect on the fluctuations of
the stress tensor operator around this mean. Similarly, the fluctuations
of distinct quantum fields should be uncorrelated, so cancellation of the
the mean values of the stress tensor does not imply any cancellation of the
fluctuations. Thus if we live in a higher dimensional world with compact
extra dimensions, it is possible for the mean stress tensor of quantum
fields in our four dimensional subspace to be zero, but for there to be large
stress tensor fluctuations around that mean which will produce large
passive metric fluctuations. These metric fluctuations could in turn
produce observable effects, such as the blurring of the images of distant 
objects. Fluctuations of the stress tensor induce Ricci curvature fluctuations,
which in turn cause the expansion parameter of a bundle of geodesics
to undergo fluctuations. This was discussed in Ref.~\cite{BF}, henceforth I, 
where the Raychaudhuri equation was treated as a Langevin equation. 
It was shown how the dispersion in the expansion $\theta$ may be calculated
as an integral of the Ricci tensor correlation function. It was also argued that
the product $s\, \theta_{rms}$  is of the order of the expected angular 
blurring and of the fractional
luminosity fluctuations of the source. Here $\theta_{rms}$ is the
 root-mean-square value of $\theta$ and $s$ is the distance to
a source. Thus sufficiently large stress tensor fluctuations would have 
already been observed and can be constrained by observation. The purpose of 
this note is to explore the extent to which this can be used to constrain
theories with compact extra dimensions.
 A possibility of observable effects
arising from the active fluctuations of the quantized gravitational field
was discussed by Yu and Ford~\cite{YF}.

\section{Compact Extra Dimensions}

Consider a flat spacetime with $4+n$ spacetime dimensions. The derivation
of the expansion fluctuations given in I for four dimensional spacetime
is essentially unchanged, apart from the fact that the Newton's constant that
appears in the Einstein equations is that for $4+n$ dimensions, $G_{4+n}$.
Consider a null geodesic with tangent vector $k^\mu$, and let $f(x)$ be
a sampling function which describes a world-tube centered on that geodesic.
Then the mean squared expansion fluctuation is
\begin{equation}
\langle (\Delta \theta)^2 \rangle =
\int d^{4+n} x \int d^{4+n} x' \, f(x) \, f(x') \,
 C_{\mu \nu \alpha \beta}(x,x')\, k^\mu (x) k^\nu (x)
\, k^\alpha(x') k^\beta(x') \,  \label{eq:ST_av}
\end{equation}
where $C_{\mu \nu \alpha \beta}(x,x')$ is the Ricci tensor correlation 
function. We take $k^\mu$ to be a constant vector.
The contraction of the correlation function into this tangent vector is
\begin{equation}
C_{\mu \nu \alpha \beta}(x,x')\, k^\mu k^\nu k^\alpha k^\beta =
128\, \pi^2\, G_{4+n}^2\,(k^\mu \Delta x_\mu)^4 \, (D'')^2 \, ,
\end{equation}
where $D$ is the the appropriate two-point function.

Here we are interested in the vacuum state for a spacetime with periodic
compactification in each of the $n$ extra space dimensions. However, let
us first consider uncompactified $4+n$ dimensional Minkowski spacetime, for
which the vacuum two point function is 
\begin{equation}
D_0(x-x') = \frac{\Gamma(\frac{n}{2} +1)}{2\,(2 \pi)^{\frac{n}{2} +2}}
\; \frac{1}{\sigma^{\frac{n}{2} +1}} \,.
\end{equation}
Here
\begin{equation}
\sigma = \frac{1}{2}\left(-\Delta t^2 +\Delta x^2 +\Delta y^2 
+\Delta z^2 +  \sum_{j=1}^n \Delta w_j^2 \right)
\end{equation}
is the invariant interval and $\Delta w_j$ is the coordinate separation in 
the $j$-th extra dimension. 

We now compactify this spacetime with periodicity lengths $L_j$ in the 
$j$-th extra dimension. The vacuum two point function now becomes
\begin{equation}
D(x-x') = \frac{\Gamma(\frac{n}{2} +1)}{2\,(2 \pi)^{\frac{n}{2} +2}}
\sum_{m_1=-\infty}^\infty \cdots \sum_{m_n=-\infty}^\infty
\; \frac{1}{\sigma_{m_j}^{{n}/{2} +1}} \,,
\end{equation}
where
\begin{equation}
\sigma_{m_j} = 
\frac{1}{2}\left[-\Delta t^2 +\Delta x^2 +\Delta y^2 +\Delta z^2 +
         \sum_{j=1}^n (\Delta w_j +m_j\,L_j)^2 \right]
\end{equation} 
and $m_j$ is the winding number in the $j$-th extra dimension. The
quantity $D''$ is formed by taking the second derivative of $D$ with
respect to $\sigma_{m_j}$ inside the summation:

\begin{equation}
D'' = \frac{\Gamma(\frac{n}{2} +3)}{2\,(2 \pi)^{\frac{n}{2} +2}}
\sum_{m_1 =-\infty}^\infty \cdots \sum_{m_n=-\infty}^\infty
\; \frac{1}{\sigma_{m_j}^{{n}/{2} +3}} \,.
\end{equation}

We are primarily interested in the limit in which the compactification lengths
are all small compared to the length scales which characterize the bundle of
rays in the uncompactified dimensions. In this case, we can replace the 
summations by integrations over continuous variables:
\begin{equation}
\sum_{m_j= -\infty}^\infty \rightarrow  \int_{-\infty}^\infty d m_j
 = \frac{1}{L_j}\, \int_{-\infty}^\infty d \xi_j \,,
\end{equation}
where $\xi_j = \Delta w_j +m_j\,L_j$. In this limit, we can write
\begin{equation}
D'' = \frac{\Gamma(\frac{n}{2} +3)\,2^{\frac{n}{2} +3}}
{2\,(2 \pi)^{\frac{n}{2} +2}\, V_C}\, S_n\, 
\int_0^\infty \frac{d\rho \, \rho^{n-1}}{(\rho^2 +r^2 -uv)^{\frac{n}{2} +3}}\, .
                                          \label{eq:DPP}
\end{equation}
Here
\begin{equation}
V_C = \prod_{j=1}^n L_j
\end{equation}
is the volume of the compactified subspace, $u = \Delta t - \Delta x$,
 $v = \Delta t + \Delta x$, $\rho^2 = \sum_{j=1}^n \xi_j^2$, and $r$
is the distance from the central line of the bundle. In addition,
we have used 
\begin{equation}
\int_{-\infty}^\infty \prod_{j=1}^n d\xi_j 
                              = S_n \int_0^\infty d\rho\, \rho^{n-1} \,,
\end{equation}
where 
\begin{equation}
S_n = \frac{2\, \pi^\frac{n}{2} }{\Gamma(\frac{n}{2})} \,.
\end{equation}
The integral in Eq~(\ref{eq:DPP}) may be explicitly evaluated to yield
\begin{equation}
D'' = \frac{2}{\pi^2\, V_C\, (r^2 -uv)^3} \,.
\end{equation}
Note that the explicit dependence upon $n$, the number of compact dimensions
drops out, apart from the factor of $V_C$.

We can now follow the same procedure as used in I to find the contribution
of the pure vacuum term in four dimensional Minkowski spacetime, with
the result
\begin{equation}
\langle (\Delta \theta)^2 \rangle = 
\frac{256 \, a^2 \,G_{4+n}^2}{5 \pi^2\, V_C^2}\; 
\left\langle \frac{1}{r^8} \right\rangle \,. \label{eq:theta_higher}
\end{equation}
The quantity $\langle {1}/{r^8} \rangle$ is an average taken over the 
bundle of rays. It can be defined by an integration by parts procedure,
and will have a value determined by the geoemtric parameters of the bundle.
In Kaluza-Klein theories, the Newton's constant in higher
dimensions, $G_{4+n}$, is related to the effective Newton's constant
in four dimensions by
\begin{equation}
G_{4+n} = V_C\, G_{4}\,.  \label{eq:Newton}
\end{equation}
This relation follows from the Einstein action in $4+n$ dimensions,
and an assumption that we can trivially integrate over the extra
dimension to obtain the effective four-dimensional action:
\begin{equation}
S = \frac{1}{16\, \pi\,G_{4+n}} \int d^{4+n} x \,\sqrt{-g}\, R
 = \frac{V_C}{16\, \pi\,G_{4+n}} x \int d^{4} x \,\sqrt{-g}\, R
 = \frac{1}{16\, \pi\,G_{4}} \int d^{4} x \,\sqrt{-g}\, R\,.
\end{equation}
However, if we use the relation between $G_{4+n}$ and $G_{4}$,
the dependence upon $V_C$ cancels, and we have exactly the same
result as for four dimensional Minkowski spacetime in the vacuum state:
\begin{equation}
\langle (\Delta \theta)^2 \rangle = 
\frac{256 \, a^2 \,\ell_P^2}{5 \pi^2\, }\; 
\left\langle \frac{1}{r^8} \right\rangle \,.  \label{eq:theta_four}
\end{equation}
Because $\langle {1}/{r^8} \rangle$ is of order $1/b^8$, where $b$ is the
transverse dimension of the bundle of rays, this quantity is very 
small~\cite{BF}. 

\section{Discussion}

We argued that quantum fluctuations for a system confined in a small
spatial volume should grow as the volume decreases. This is indeed
reflected in Eq.~(\ref{eq:theta_higher}), where 
$\langle (\Delta \theta)^2 \rangle \rightarrow \infty$ as $V_C \rightarrow 0$
for fixed $G_{4+n}$. The special feature of Kaluza-Klein theories
which permits them to avoid exhibiting large focusing fluctuations in
the four-dimensional world is the relation Eq.~(\ref{eq:Newton}). If we
let $V_C \rightarrow 0$ with fixed $G_{4}$, then the gravitational coupling
vanishes at just the rate needed to mask the effects of the extra
dimensions. Thus one cannot place any constraints on Kaluza-Klein theories
from phenomena such as focusing fluctuations which scale as powers of
$G_{4+n}/V_C$.

\begin{acknowledgments}
  This work was supported in part by the National
Science Foundation under Grant PHY-0244898.
\end{acknowledgments}

\end{document}